\documentclass{IOS-Book-Article}

\usepackage{mathptmx}
\usepackage{graphicx}
\usepackage{amssymb}
\usepackage{amsmath}
\usepackage{aas_macros}
\usepackage{color}

\def\ene{\epsilon}

\begin{document}
\begin{frontmatter}              % The preamble begins here.

%\pretitle{Pretitle}
\title{Towards Petaflops Capability of the VERTEX Supernova Code}
\runningtitle{Towards Petaflops Capability of the VERTEX Supernova Code}
%\subtitle{Subtitle}

\author[A]{\fnms{Andreas} \snm{Marek}\thanks{Corresponding Author:
    Andreas Marek; E-mail: andreas.marek@rzg.mpg.de.}%
},
\author[A]{\fnms{Markus} \snm{Rampp}},
\author[B]{\fnms{Florian} \snm{Hanke}},
and
\author[B]{\fnms{Hans-Thomas} \snm{Janka}}

\runningauthor{A. Marek et al.}
\address[A]{Computing Centre (RZG) of the Max-Planck-Society\\and the
Max-Planck-Institute for Plasma Physics\\ Boltzmannstrasse 2, 85748 Garching, Germany.}
\address[B]{Max-Planck-Institute for Astrophysics\\
Karl-Schwarzschildstr. 1, 85748 Garching, Germany.}

\begin{abstract}
The VERTEX code is employed for multi-dimensional 
neutrino-radiation hydrodynamics simulations of core-collapse supernova
explosions from first principles. The code is considered state-of-the-art in supernova
research and it has been used for modeling 
for more than a decade, resulting in numerous scientific publications.
The computational performance of the code, which is currently deployed 
on several high-performance computing (HPC) systems up to the Tier-0
class (e.g.\ in the framework of the European PRACE initiative and the 
German GAUSS program), however, has so far not been extensively documented.
This paper presents a high-level overview of the relevant algorithms and
parallelization strategies and outlines the 
technical challenges and achievements encountered along the evolution of the code 
from the gigaflops scale with the first, serial simulations in 2000, 
up to almost petaflops capabilities, as demonstrated 
lately on the SuperMUC system of the Leibniz Supercomputing Centre
(LRZ).
In particular, we shall document the parallel scalability and 
computational efficiency of VERTEX at the large scale and on the
major, contemporary HPC platforms. We will outline upcoming
scientific requirements and discuss the resulting challenges for the
future development and operation of the code.
\end{abstract}

\end{frontmatter}

\thispagestyle{empty}
\pagestyle{empty}

\paragraph{Acknowledgments}
A.\ Marek and M.\ Rampp are grateful for fruitful 
discussions with the staff of IBM, especially with Florian Merz and Francois Thomas.
The authors thank the J\"ulich Supercomputing Centre (JSC), the
Leibniz Supercomputing Centre (LRZ), the High Performance Computing
Center Stuttgart (HLRS) and the CEA in France for supporting large-scale 
benchmarks with VERTEX on their HPC systems. At the Max-Planck-Institute for Astrophysics, research with 
VERTEX-PROMETHEUS is supported by the Deutsche Forschungsgemeinschaft
through grants SFB/TR-7 and EXC-153. High-performance computing
resources (Tier-0) are provided by PRACE on CURIE TN (GENCI@CEA,
France) and SuperMUC (GCS@LRZ, Germany) and by the Gauss Centre 
for Supercomputing on SuperMUC (GCS@LRZ, Germany).

\section{Introduction}

Theoretical modeling of core-collapse supernovae, specifically
the attempt to understand the still unknown explosion mechanism from first
principles, is an extremely challenging multi-dimensional, multi-scale, 
multi-physics problem. For decades, numerical simulations of this 
spectacular astrophysical phenomenon have been at the forefront 
of computational physics and high-performance computing, usually 
pushing the limits of the supercomputers of their time. 
Unlike almost any other known (astro-)physical scenario, the weakly
interacting neutrinos released during the gravitational collapse
of a massive star play a subtle \emph{dynamical} role in the evolution and are thought to ultimately power the 
observed supernova explosion \cite{Bethe1985,Wilson1985}.
The leaking of 
neutrinos out of the extremely dense interior of the collapsed star 
occurs on timescales 
which are relevant for the overall dynamics, and their reabsorption 
in the surrounding layers happens under semi-transparent conditions. 
As a consequence, the transport of energy, momentum and lepton number
has to be treated numerically very accurately, by following the time-evolution of
the neutrino distribution function in six-dimensional phase-space, 
as governed by the Boltzmann equation. Together with the 
coupling to the evolution of the stellar material (by virtue of the exchange of 
energy, momentum and lepton number) and the dynamical evolution of the latter
this constitutes a computationally extremely expensive
radiation-hydrodynamics problem, not to forget an adequate treatment
of the microphysics and gravitation. 
Only at the beginning of this century, modeling the time-evolution 
of the phase-space distribution of neutrinos at the level of
the Boltzmann equation and the coupling to the evolution of
the stellar material has become computationally tractable at all.
But even with the assumption of spherical symmetry of the stellar medium,
which is a severe approximation, a three-dimensional and
time-dependent transport problem has to be solved which already stressed 
the supercomputers of the early 2000's to their limits
\cite{Rampp2000,Rampp2000a,Burrows2000,Liebendoerfer2001c,Liebendorfer2001d}. Meanwhile, due 
to the ever-increasing computing power, but also thanks to algorithmic 
and technical innovations it has become possible to treat the full 
three-dimensional evolution of the stellar material coupled to 
increasingly accurate neutrino transport \cite[and references cited
  therein]{Janka2012_review,Cardall2012}. For the latter, a 
few --- apparently reasonable --- approximations are commonly adopted
which effectively reduce the dimensionality of the transport problem, 
as computing genuinely six-dimensional, time-dependent transport solutions still appears 
out-of-reach in this context.   

VERTEX (\textbf{V}ariable \textbf{E}ddington Factor \textbf{R}adiation \textbf{T}ransport for Supernova
\textbf{Ex}plosions) is one of the very few simulation codes at this level of
physical accuracy and comprehensiveness. The code has been around since 
2000, has been continuously upgraded by new physics and new algorithmic features
   since then, and is considered state-of-the-art in the
   field. Simulations with VERTEX  
have played a decisive role for establishing the view that the explosion mechanism of 
core-collapse supernovae is a genuinely multi-dimensional 
phenomenon \cite{Rampp2000a,Buras2003} and the code has been spear-heading multi-dimensional 
modeling since then. While the algorithms, their implementation in 
the code and its verification have been extensively documented \cite{Rampp2000,Rampp2002b,Liebendoerfer2005,Marek2006a}, 
and numerous physics papers have been produced \cite[and references
  cited therein]{Janka2012_review}, the computational 
performance of the code is not yet published in much detail. 

\smallskip

Since the start of its development in the late 1990s, and the first 
simulation runs \cite{Rampp2000,Rampp2000a} on a single CPU of a NEC SX-5 vector system
(gigaflops-performance scale), the VERTEX code has been continuously ported to, and 
used in production on all major HPC 
platforms (IBM Bluegene and Power, CRAY, x86, x86\_64), in particular
at the three national German HPC centers, the computing center of the Max-Planck-Society (RZG), and
various Tier-0 systems of the European PRACE infrastructure.
During the last 15 years, and a man-power equivalent of almost half a century,
beginning with a serial code and spherically-symmetric models, 
VERTEX has evolved to a highly tuned, hybrid MPI$/$OpenMP-parallelized HPC
code for multi-dimensional simulations of core collapse supernovae 
\cite{Buras2006a,Buras2006b,Marek2009a,Hanke2013a}.
Today, the code is typically operated in production at 10\dots
100~teraflops, using tens of thousands of cores (x86\_64). Very recently, 
at the "SuperMUC Extreme Scaling Workshop, 2013" of the Leibniz Supercomputing Centre (LRZ), we 
have demonstrated close-to-petascale performance, using all 131\,000 cores
of the SuperMUC system. 
Importantly, benchmarks at that scale are not mere showcases, as  
scientific progress with supernova modeling is still heavily computing-time limited,
and there is a high demand for further optimization and parallel
scaling beyond sustained petascale. Specifically, there is a strong \emph{science-driven need} \cite{Janka2012_talk}
for typical simulation runs to get even bigger,
i.e.\ more highly resolved (partly addressable by weak scaling) and 
significantly faster, i.e.\ concerning the time to solution of an individual
model run which currently takes many weeks (addressable by strong
scaling). 

\smallskip

This paper is organized as follows: Section~\ref{sect:math_model} provides a high-level overview
of the algorithm and the parallelization approach adopted for
VERTEX. Section~\ref{sect:performance} describes its
computational performance in terms of parallel scalability as well
as absolute floating-point performance. The subsequent
Section~\ref{sect:developments} outlines our ongoing and 
future developments before we conclude in Section~\ref{sect:conclusions}.

\section{High-level description of the algorithm}
\label{sect:math_model}

\subsection{Mathematical and numerical model}
The VERTEX neutrino-transport code \cite{Rampp2002b} is coupled to the hydrodynamics 
code PROMETHEUS \cite{Fryxell1989}, which is a dimensionally split 
implementation of the piecewise parabolic method (PPM \cite{Colella1984}).
This specific combination, termed PROMETHEUS-VERTEX, is the one that has been
employed most commonly so far, but VERTEX has been successfully 
operated with alternative hydrodynamics solvers as well \cite{Mueller2010,Mueller2010b}.
We note that the distinction between different hydrodynamics solvers is 
not very relevant in the present context, as the basic parallelization 
approach and the performance characteristics are largely dominated by 
the neutrino-transport module. Hence, similiar performance figures
could in principle be reached when VERTEX is operated together with other
hydrodynamics solvers like, e.g. in the context of the COCONUT-VERTEX 
code \cite{Mueller2010,Mueller2010b}.  

The non-linear system of partial differential equations which is 
solved in PROMETHEUS-VERTEX consists of the following components:
\begin{itemize}
\item
The Euler equations of hydrodynamics,
supplemented by advection equations for the electron fraction and the
chemical composition of the fluid, and formulated in spherical
coordinates;
\item
the Poisson equation for calculating the gravitational source terms
which enter the Euler equations, including corrections for general
relativistic effects;
\item
the Boltzmann transport equation which determines the
(non-equilibrium) distribution function of the neutrinos;
\item
the emission, absorption, and scattering rates of neutrinos, which
are required for the solution of the Boltzmann equation;
\item
the equation of state of the stellar fluid, which provides the closure
relation between the variables entering the Euler equations, i.e.
density, momentum, energy, electron fraction, composition, and pressure.
\end{itemize}

\noindent
The neutrino transport is the numerically and computationally most challenging
part, thus we will briefly summarize the relevant algorithms here. For a more complete description of the entire code we
refer the reader to \cite{Buras2006a}, and the references therein.

\begin{figure}[!h]
\centering
\includegraphics[width=5cm]{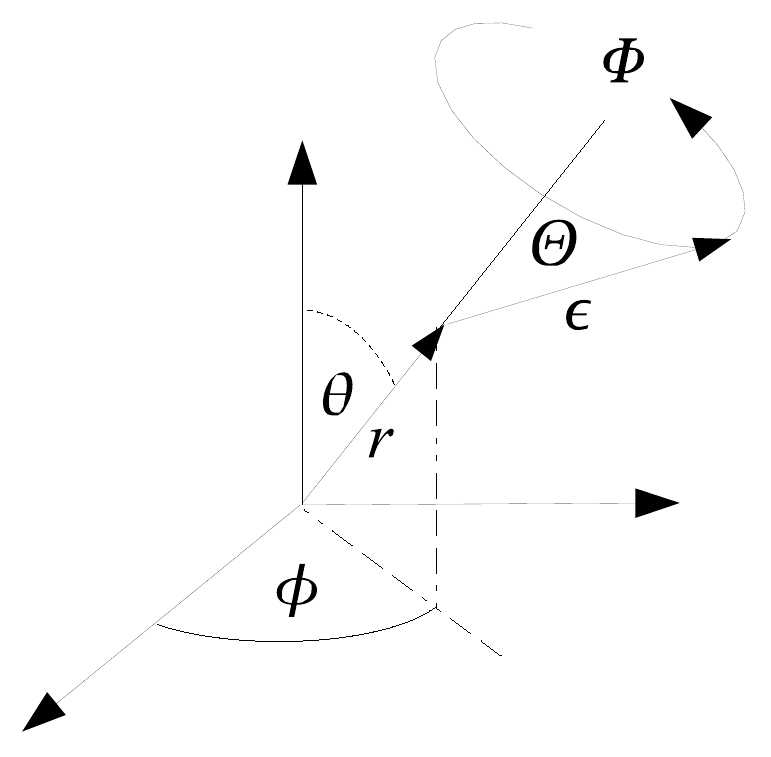}
\caption{Illustration of the phase space coordinates (see the main text).}
\label{fig:coords}
\end{figure}

The crucial quantity required to determine the source terms for the
energy, momentum, and electron fraction of the fluid owing to its
interaction with the neutrinos is the neutrino distribution function
in phase space $f(r,\theta,\phi,\epsilon,\Theta,\Phi,t)$. More often
the neutrino intensity $I = c/(2 \pi \hbar c)^3 \cdot \epsilon^3 f $
is used.

It describes, at
every point in space $(r,\theta,\phi)$, the phase-space density of
neutrinos propagating with energy $\epsilon$ into the direction
$(\Theta,\Phi)$ at time $t$ (see Fig.~\ref{fig:coords}).

The evolution of the phase-space density $f$ in time, or equivalently of the neutrino intensity $I = c/(2 \pi \hbar c)^3 \cdot \epsilon^3 f $, is governed by the
Boltzmann equation, and solving this equation is, in general, a
six-dimensional problem (as time is usually not counted as a
separate dimension) and is hampered by the fact that the scattering
terms in the interaction operator constitute an integro-differential problem. 

The dimensionality of the problem can be reduced by forming angular 
moments of $I$ (via the integration over momentum space) and solving 
evolution equations for these moments. The 0$^\mathrm{th}$ to
3$^\mathrm{rd}$-order moments are defined as
\begin{equation}
J,\boldsymbol{H},{\mathit{K}},{\mathit{L}},
\dots (r,\theta,\phi,\ene,t) = \frac{1}{4 \pi}
   \int I (r,\theta,\phi,\ene,\Theta,\Phi,t) \,
                  \boldsymbol{n}^{0,1,2,3,\dots} \,{\rm d} \Omega ,
\end{equation}
where $ {\rm d} \Omega = \sin \Theta \, {\rm d} \Theta \, {\rm
d}\Phi$, $\boldsymbol{n} = (\cos \Theta,\sin \Theta \cos \Phi,\sin
\Theta \sin \Phi )$, and exponentiation represents repeated
application of the dyadic product. Note that the moments are
\emph{tensors} of the respective rank.

It can be shown that in order to compute the source
terms for the energy, momentum and electron fraction of the fluid, it is
sufficient to solve the transport equations for the zeroth moment $J$ 
(neutrino energy density) and first moment $H$ (neutrino flux) of the
relativistic, comoving-frame Boltzmann transport equation in the 
$\mathcal{O}(v/c)$ approximation \cite{Mihalas1984}. This set of equations can be closed
by so-called variable Eddington factors \cite{Mihalas1984} , which in
our case are derived from the solution of a separate, simplified
``model'' Boltzmann equation.

A finite volume discretization of the moment equations is solved on the
spatial domain
$[0,r_\text{max}]\times[\theta_\text{min},\theta_\text{max}]\times[\phi_\text{min},\phi_\text{max}]$,
where $\theta_\text{min}=0$ and $\theta_\text{max}=\pi$ correspond to
the north and south poles, respectively, of the spherical grid.

The equations are solved in three operator-split steps, corresponding to
a lateral, azimuthal, and a radial sweep. The angular sweeps describe
the advection of neutrinos with the stellar fluid, and thus couple the
angular moments of the neutrino distribution of neighbouring angular
zones. They are computed with an \emph{explicit} upwind scheme.
The radial sweeps are solved \emph{implicitly} in time\footnote{Explicit schemes
would enforce very small time steps to cope with the stiffness of the
source terms in the optically thick regime, and the small CFL time step dictated by
neutrino propagation with the speed of light in the optically thin
regime. Still, even with an implicit scheme $\gtrsim 10^6$ time steps
are required per simulation, rendering the calculations very
expensive.}. The radial sweeps are performed \emph{"ray-by-ray"},
i.e.\ separately for each angular direction $(\theta,\phi)$, employing
a second-order scheme with backward differencing in time. For each
"ray" this leads to a non-linear
system of algebraic equations, which is solved by Newton-Raphson
iteration with explicit construction of the Jacobian and direct
solution of the linearized system. The ray-by-ray ansatz assumes that
the phase space distribution function is axially symmetric around the 
radial direction. This approach implies that the neutrino fluxes are 
purely radial, which is a reasonably good approximation if
nonspherical asymmetry of the neutron star as a neutrino source does not play a role.

\subsection{Parallelization strategy}\label{sect:parallel_strategy}

By virtue of the the ''ray-by-ray'' approximation outlined above, the
three spatial dimensions can be decomposed, leading to a two-dimensional set 
in $(\theta,\phi)$-space of one-dimensional ($r$, plus two
phase-space dimensions, $\ene$ and $\Theta$) problems which are
only loosely coupled and can be computed independently.
The parallelization takes advantage of this scheme by adopting a 
classical, two-dimensional decomposition of the spherical grid
into angular MPI-domains. This is complemented by a coarse-grained OpenMP 
parallelization over the rays $(\theta,\phi)$ within the
MPI-domain, leading to a many-to-one mapping of rays to threads,
i.e.\ at most one thread (or CPU core) works on a single ray.
 
Apart from a few global reductions (e.g. in the computation of the gravitational potential, or for
determining time-step limits) for which the collective
\texttt{MPI\_ALLREDUCE} operation is used,
information between individual rays needs to be exchanged only 
among nearest spatial neighbours. 
In effect, our hybrid parallelization approach decreases the surface-to-volume ratio of
the individual MPI domains (with respect to a "flat" MPI
parallelization). Thus, the communication volume which is proportional
to the surface of an MPI domain in coordinate space can be packed into a 
smaller number of larger messages (exchanged using \texttt{MPI\_SEND} and 
\texttt{MPI\_RECV}), which reduces the latency
penalty of the network and increases bandwith usage.

\section{Computational performance}\label{sect:performance}
\subsection{Scaling results}\label{sect:scaling_results}

 \begin{figure*}[!h]
 \resizebox{0.75\linewidth}{!}{\includegraphics{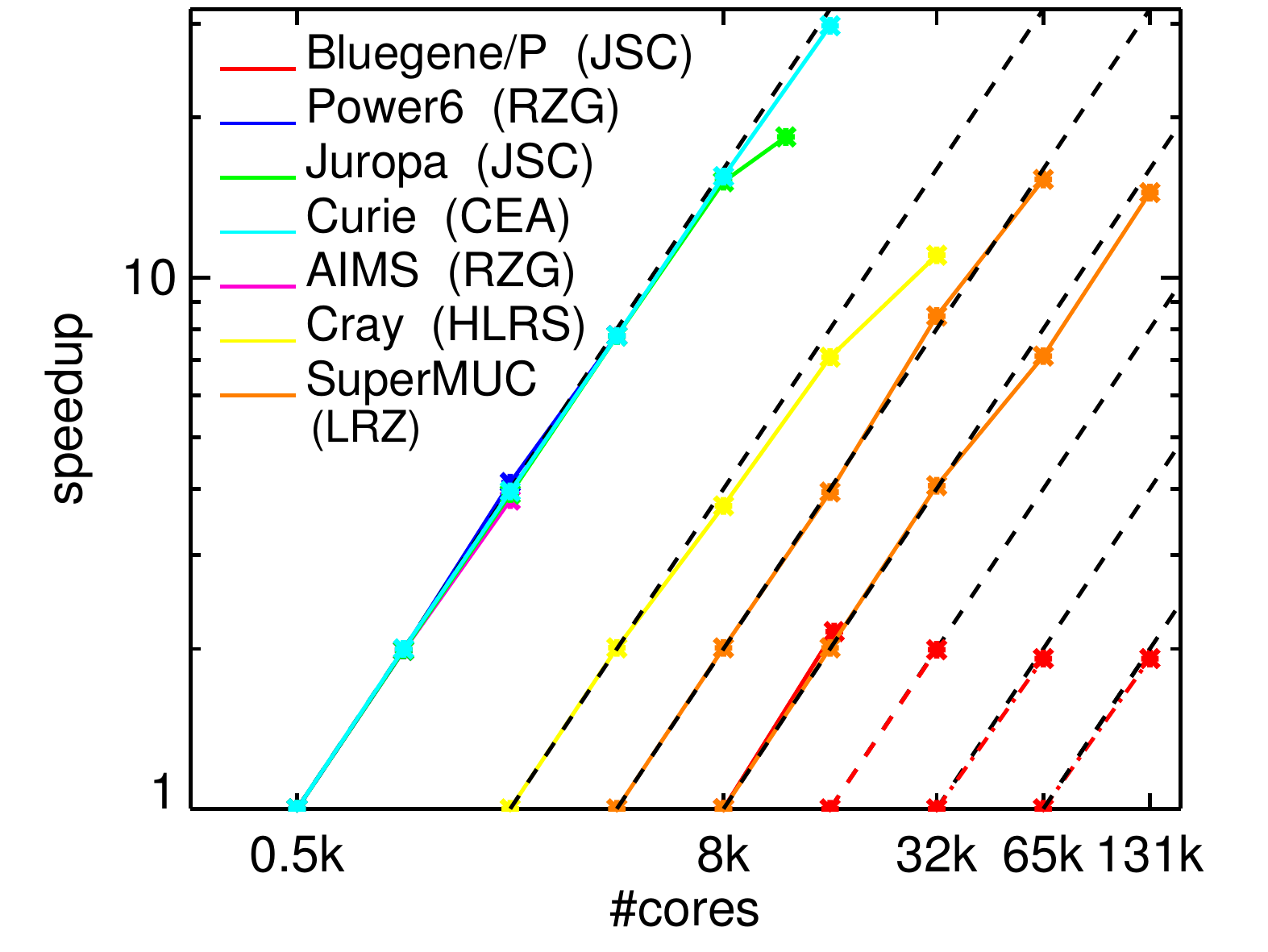}}
 \caption{Strong scaling of PROMETHEUS-VERTEX on different machines, namely on
   clusters with x86\_64 processors and high-performance
   interconnects (Juropa, Aims, with Intel Nehalem CPUs; Curie,
   SuperMuc with Intel SandyBridge CPUs), on an IBM Power6 and Bluegene/P,
   and on a Cray XE6. On Bluegene/P and SuperMUC it is --- due to memory requirements --- not possible to scale the
   code all the way from 8k to 131k, or from 4k to 131k, respectively. Thus, several
   scaling runs with adapted model sizes were done to bridge the gap:
   on Bluegene/P from 8k to 16k, from 16k to 32k, from 32k to 65k, and from 65k to 131k cores, 
   respectively. On SuperMUC,
   scaling runs were done from 4k to 65k, and from 8k to 131k cores,
   respectively. Dashed lines indicate the ideal speedup. The dips
   in the speedup curves for Juropa and Cray are understood and the
   causes are resolved. However, there was no opportunity to rerun the
   benchmarks on these machines.}\label{scaling}
\end{figure*}

VERTEX has been demonstrated to deliver excellent weak (over the
number of angular rays) and strong scalability (over a wide range of core counts,
provided there are more angular rays than available cores).
In Fig.~\ref{scaling} we show the strong-scaling behavior of the code for
different HPC machines and architectures over the last couple of years. The
strong-scalability is almost perfect\footnote{The breakdowns
on 12\,000 cores on JUROPA and on 32\,000 cores on the
Cray of HLRS could meanwhile be explained and are already fixed.}. 
Typical production runs are currently performed using 16\,000 cores, where
the scaling is perfect and parallel efficiency is excellent.

On SuperMUC, as well as on Bluegene/P very good scaling holds up to 
131\,000 cores. This excellent scaling behaviour is
in the first place achieved by design of the
domain-decomposition into loosely coupled rays, with additional latency 
and bandwidth optimization by the hybrid MPI/OpenMP parallelization. 
Thus, the time spent in communication is at most on the 10\% level. 
Secondly, the memory layout of the code has been ''localized'' on the
node level as well as on the entire computational domain. Except for a few
arrays, which are necessary for basic book keeping, the MPI domains do
not require any knowledge of each other, which leads to a scalable memory 
footprint. On the node level, it is crucial in our approach to
achieve a next-to-perfect parallel efficiency of the OpenMP
parallelization over rays within an MPI domain. Optimizing data 
locality on ccNUMA architectures turned out to be an important
measure.

\subsection{Floating-point performance}

We analyse the floating-point performance of PROMETHEUS-VERTEX within the roofline model
\cite{Williams2009}, which allows to classify the performance of an
application and helps to identify and visualize its fundamental
hardware and algorithm-specific bounds in the two-dimensional space of 
memory-bandwidth and compute-performance coordinates. 
\begin{figure*}[!h]
 \resizebox{0.75\linewidth}{!}{\includegraphics{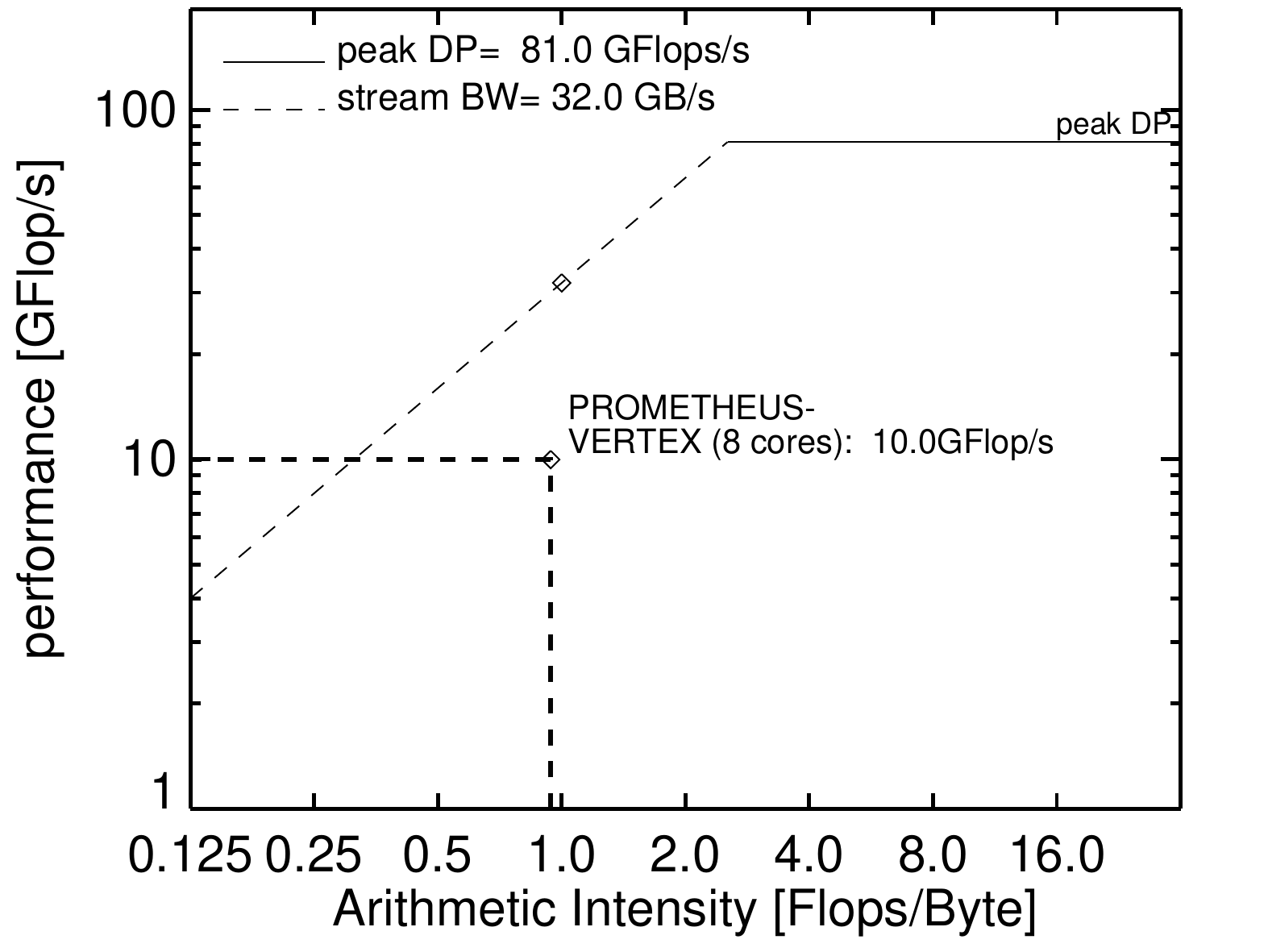}}
 \caption{Roofline model for an eight-core node with two Intel
Xeon E5540 "Nehalem" CPUs (2.53 GHz), assuming double-precision calculations. The solid
   horizontal line shows the nominal peak performance of the node, which is 81 GFlop/s. The
   dashed inclined line is given by the memory bandwidth (32 GB/s) as measured with
   the stream benchmark \cite{stream}. The maximum achievable performance of an
   application is bounded to the region below these "roof"-lines. PROMETHEUS-VERTEX (symbol) has an algorithmic intensity of 0.96 and achieves 10 GFlop/s.}\label{fig:vertex_roofline}
\end{figure*}
The analysis was done on an eight-core compute node with two Intel
Xeon E5540 "Nehalem" CPUs\footnote{Due to known
  problems with the hardware counters on the current Intel SandyBridge 
  microarchitecture \cite{sandybridge_flops}, measurements of the
  floating-point performance on such CPUs have to be interpreted with caution.} with a frequency of
2.53 GHz.  Measurements with three
independent tools, namely \texttt{likwid} \cite{likwid},
\texttt{papi} \cite{papi}, and \texttt{perflib} \cite{perflib}, gave consistent results.

Figure~\ref{fig:vertex_roofline} shows the maximum of the achievable
floating-point performance as a function of the arithmetic intensity
(AI) of the algorithm, i.e.\ the number of double-precision
floating-point operations per byte transferred from main memory. These two "roof"-lines
are characteristic for the given hardware and divide the diagram into
a ''memory bound'' (AI$<2.2$) and a ''compute bound'' (AI$>2.2$) region.
With AI=0.96, PROMETHEUS-VERTEX is still memory bound on this processor, but operates with more
than 12 percent of the peak performance, or with about 30 percent of
the performance which can theoretically be reached with this arithmetic intensity. 
Since about 86\% of all instructions are already SIMD-vectorized, the
first target for further performance optimization of
the code is reducing the number of memory references, i.e. trying to shift the
arithmetic intensity towards higher values. Nevertheless for a complex scientific
application with a non-trivial instruction mix, values of around 10 percent
of the nominal peak floating-point performance are usually considered
very good.

By scaling the performance measured on the Nehalem node with the
runtime measured on a SandyBridge node (2x Xeon E5-2680 with 320 GFlop/s 
peak performance), we derive a floating-point performance
of $\approx$ 35 GFlop/s/node (11\% of the peak performance). Scaling with
the measured parallel efficiency for weak scaling from 1 to 8192 nodes
of the SuperMUC system, we obtain
a sustained performance of roughly 0.25 PFlop/s corresponding to about
10\% of the nominal peak performance of SuperMUC.

\section{Ongoing and future developments}\label{sect:developments}

Although PROMETHEUS-VERTEX shows an excellent scaling behavior and overall 
performance, there is still a high scientific motivation to 
strive for further performance improvements in the near future. 
The primary goal is reducing the time to solution of an individual
model run which currently takes many weeks.

The currently adopted parallelization approach with its excellent
scalability over angular rays is based on a many-to-one mapping of the rays
to individual threads (or cores). This has been a great
advantage so far, as the number of zones used in typical simulations 
increased along with the growing number of available cores. 
However, with HPC architectures apparently evolving towards more and
more threads (cores) with stagnating or even decreasing
single-thread performance, this strength of the transport module, VERTEX, 
is gradually turning into a weakness: The number of angular 
rays required for supernova simulations is expected to stagnate at a level of a 
few hundreds of thousands (corresponding to angular
resolutions of less than $1^\circ$) and hence thread counts beyond a level of
$\mathcal{O}(10^6)$ cannot be utilized efficiently anymore with the
current approach. Moreover, the time to solution of less finely
resolved simulations cannot be decreased by using such large numbers
of cores.

To overcome this limit, parallelism \emph{within} an
angular ray, specifically in the phase-space coordinates can be
exploited, e.g.\ by using (nested) OpenMP threads, accelerators, or alike. Early,
one-dimensional simulations which effectively operated on a single ray have
already successfully used the OpenMP approach (however without
having to bother with the subtleties of nested OpenMP) and recently,
we have successfully used GPUs for exploiting parallelism on this 
level \cite{Dannert2013}. The nested OpenMP approach would allow to use 
at least ten to twenty times more cores on traditional multi-core
architectures, and the new GPU version of VERTEX has already delivered gains 
by a factor of two in total application performance. Similar speedups
are expected from the Intel many-integrated-cores (MIC) architecture.

\section{Summary and Conclusions}\label{sect:conclusions}

We have given an overview of the supernova-simulation code PROMETHEUS-VERTEX,
focussing on the neutrino-transport module, VERTEX, and specifically
on the adopted parallelization approach and its basic performance
characteristics on large HPC platforms. At the time of this writing 
production runs using VERTEX are typically performed using a number of
16\,000 cores with excellent parallel efficiency. 
We have demonstrated that it is possible already
now to efficiently employ the code on much bigger computers. 
In the course of the "SuperMUC Extreme Scaling Workshop, 2013" of the 
Leibniz Supercomputing Centre (LRZ), we were able to use up to
131\,000 cores of the SuperMUC system, thereby maintaining very high 
parallel efficiency and floating-point performances. The code has reached 
a quarter of a petaflop (double-precision), which is equivalent 
to about 10\% of the nominal peak performance of SuperMUC.

Importantly, there is indeed a \emph{scientifically driven need} for 
performing simulations at this scale, and hence the results we have 
documented here are highly relevant benchmarks, rather than being mere
showcases. In particular, physical completeness and scientific 
relevance of the employed setups have not been sacrificed for achieving 
high performance. 
However, although theoretically, such huge production simulations 
could already now be performed with very good performance, they
are not yet feasible in practice. Firstly, the necessary throughput
can usually not be achieved when using a significant partition of
an entire HPC system, and secondly, system stability still appears to 
be a serious issue at this scale.
Nevertheless, we expect that this situation is rapidly improving
and first-principles simulations of core-collapse 
supernova explosions employing VERTEX neutrino transport will soon be routinely performed 
at the scale of hundreds of thousands of processor cores.

Finally, we have already started preparing the code for
the next generation of HPC systems which are expected to 
provide massive SIMT and SIMD parallelism on the node level, as
indicated by current GPU accelerators or many-core coprocessors.
Non-trivial work remains to be done in order for VERTEX to be
able to efficiently exploit such massive parallelism with 
relatively weak single-thread performance, but first promising
results with GPUs and the Intel Many Integrated Core (MIC) architecture 
have already been obtained.

\bibliographystyle{unsrt}
\bibliography{JabRef}

\begin{thebibliography}{10}

\bibitem{Bethe1985}
H.~A. {Bethe} and J.~R. {Wilson}.
\newblock {Revival of a stalled supernova shock by neutrino heating}.
\newblock {\em \apj}, 295:14--23, August 1985.

\bibitem{Wilson1985}
J.~R. {Wilson}.
\newblock Supernovae and post-collapse behavior.
\newblock In J.M. {Centrella}, J.M. {LeBlanc}, and R.L. Bowers, editors, {\em
  Numerical Astrophysics}, pages 422--434, Boston, 1985. Univ. Illinois,
  Urbana-Champaign US, Jones and Bartlett.

\bibitem{Rampp2000}
M.~{Rampp}.
\newblock {\em Radiation Hydrodynamics with Neutrinos: Stellar Core Collapse
  and the Explosion Mechanism of Type~II Supernovae}.
\newblock PhD thesis, Technische Universit\"at M\"unchen, 2000.

\bibitem{Rampp2000a}
M.~{Rampp} and H.-T. {Janka}.
\newblock {Spherically Symmetric Simulation with Boltzmann Neutrino Transport
  of Core Collapse and Postbounce Evolution of a 15 M$_{\odot}$ Star}.
\newblock {\em \apjl}, 539:L33--L36, August 2000.

\bibitem{Burrows2000}
A.~{Burrows}, T.~{Young}, P.~{Pinto}, R.~{Eastman}, and T.~A. {Thompson}.
\newblock {A New Algorithm for Supernova Neutrino Transport and Some
  Applications}.
\newblock {\em \apj}, 539:865--887, August 2000.

\bibitem{Liebendoerfer2001c}
M.~{Liebend{\"o}rfer}, A.~{Mezzacappa}, F.-K. {Thielemann}, O.~E. {Messer},
  W.~R. {Hix}, and S.~W. {Bruenn}.
\newblock {Probing the gravitational well: No supernova explosion in spherical
  symmetry with general relativistic Boltzmann neutrino transport}.
\newblock {\em \prd}, 63(10):103004--+, May 2001.

\bibitem{Liebendorfer2001d}
M.~{Liebend{\"o}rfer}, A.~{Mezzacappa}, and F.-K. {Thielemann}.
\newblock {Conservative general relativistic radiation hydrodynamics in
  spherical symmetry and comoving coordinates}.
\newblock {\em \prd}, 63(10):104003--+, May 2001.

\bibitem{Janka2012_review}
H.-Th. Janka.
\newblock Explosion mechanisms of core-collapse supernovae.
\newblock {\em Annual Review of Nuclear and Particle Science}, 62:407, 2012.

\bibitem{Cardall2012}
C.~Cardall, E.~Endeve, R.~D. Budiardja, P.~Marronetti, and A.~Mezzacappa.
\newblock Towards the core-collapse supernova explosion mechanism.
\newblock In {\em Advances in computational astrophysics: methods, tools, and
  outcome}, volume 453 of {\em ASP Conference Proceedings}, page~81, 2012.

\bibitem{Buras2003}
R.~{Buras}, M.~{Rampp}, H.-Th. {Janka}, and K.~{Kifonidis}.
\newblock {Improved Models of Stellar Core Collapse and Still No Explosions:
  What Is Missing?}
\newblock {\em Physical Review Letters}, 90(24):241101--+, June 2003.

\bibitem{Rampp2002b}
M.~{Rampp} and H.-Th. {Janka}.
\newblock Radiation hydrodynamics with neutrinos. variable eddington factor
  method for core--collapse supernova simulations.
\newblock {\em \aap}, 396:361--392, December 2002.

\bibitem{Liebendoerfer2005}
M.~{Liebend{\"o}rfer}, M.~{Rampp}, H.-T. {Janka}, and A.~{Mezzacappa}.
\newblock {Supernova Simulations with Boltzmann Neutrino Transport: A
  Comparison of Methods}.
\newblock {\em \apj}, 620:840--860, February 2005.

\bibitem{Marek2006a}
A.~{Marek}, H.~{Dimmelmeier}, H.-Th. {Janka}, E.~{M{\"u}ller}, and R.~{Buras}.
\newblock Exploring the relativistic regime with {Newtonian} hydrodynamics: an
  improved effective gravitational potential for supernova simulations.
\newblock {\em \aap}, 445:273--289, January 2006.

\bibitem{Buras2006a}
R.~{Buras}, M.~{Rampp}, H.-T. {Janka}, and K.~{Kifonidis}.
\newblock {Two-dimensional hydrodynamic core-collapse supernova simulations
  with spectral neutrino transport. I. Numerical method and results for a 15
  M{\.o} star}.
\newblock {\em \aap}, 447:1049--1092, March 2006.

\bibitem{Buras2006b}
R.~{Buras}, H.-T. {Janka}, M.~{Rampp}, and K.~{Kifonidis}.
\newblock {Two-dimensional hydrodynamic core-collapse supernova simulations
  with spectral neutrino transport. II. Models for different progenitor stars}.
\newblock {\em \aap}, 457:281--308, October 2006.

\bibitem{Marek2009a}
A.~{Marek} and H.-T. {Janka}.
\newblock {Delayed Neutrino-Driven Supernova Explosions Aided by the Standing
  Accretion-Shock Instability}.
\newblock {\em \apj}, 694:664--696, March 2009.

\bibitem{Hanke2013a}
F.~{Hanke}, B.~{M{\"u}ller}, A.~{Wongwathanarat}, A.~{Marek}, and H.-T.
  {Janka}.
\newblock {SASI Activity in Three-dimensional Neutrino-hydrodynamics
  Simulations of Supernova Cores}.
\newblock {\em \apj}, 770:66, June 2013.

\bibitem{Janka2012_talk}
H.-Th. Janka.
\newblock Explosion models of core-collapse supernovae.
\newblock
  http://theorie.ikp.physik.tu-darmstadt.de/nhc/pages/events/hirschegg/2013/talks/Wed/Janka.pdf,
  2012.
\newblock Hirschegg.

\bibitem{Fryxell1989}
B.A. {Fryxell}, E.~{M\"uller}, and W.D. {Arnett}.
\newblock Hydrodynamics and nuclear burning.
\newblock preprint MPA--449, Max Planck Institut f\"ur Astrophysik, Garching,
  apr 1989.

\bibitem{Colella1984}
P.~{Colella} and P.~R. {Woodward}.
\newblock {The Piecewise Parabolic Method (PPM) for Gas-Dynamical Simulations}.
\newblock {\em Journal of Computational Physics}, 54:174--201, September 1984.

\bibitem{Mueller2010}
B.~{M{\"u}ller}, {H.-T.} {Janka}, and H.~{Dimmelmeier}.
\newblock {A New Multi-dimensional General Relativistic Neutrino Hydrodynamic
  Code for Core-collapse Supernovae. I. Method and Code Tests in Spherical
  Symmetry}.
\newblock {\em \apjs}, 189:104--133, July 2010.

\bibitem{Mueller2010b}
B.~{M{\"u}ller}, H.-T. {Janka}, and A.~{Marek}.
\newblock A new multi-dimensional general relativistic neutrino hydrodynamics
  code for core-collapse supernovae. ii. relativistic explosion models of
  core-collapse supernovae.
\newblock {\em The Astrophysical Journal}, 756(1):84, 2012.

\bibitem{Mihalas1984}
D.~{Mihalas} and B.W. {Mihalas}.
\newblock {\em Foundations of Radiation Hydrodynamics}.
\newblock Oxford University Press, 1984.

\bibitem{Williams2009}
Samuel Williams, Andrew Waterman, and David Patterson.
\newblock Roofline: an insightful visual performance model for multicore
  architectures.
\newblock {\em Commun. ACM}, 52(4):65--76, April 2009.

\bibitem{stream}
{STREAM}.
\newblock Sustainable memory bandwidth in high performance computers.
\newblock http://www.cs.virginia.edu/stream/.

\bibitem{sandybridge_flops}
Likwid Forum.
\newblock
  http://likwid-tools.blogspot.de/2012/02/intel-sandybridge-and-counting-flops.html.

\bibitem{likwid}
J.~{Treibig}, G.~{Hager}, and G.~{Wellein}.
\newblock {LIKWID: Lightweight Performance Tools}.
\newblock {\em ArXiv e-print 1104.4874}, April 2011.

\bibitem{papi}
PAPI.
\newblock Performance application programming interface.
\newblock http://icl.cs.utk.edu/papi.

\bibitem{perflib}
R.~Tisma and T.~Dannert.
\newblock Perflib: A simple performance library, {RZG}.
\newblock
  http://www.rzg.mpg.de/services/computing/software/performance-tools/perf.html.

\bibitem{Dannert2013}
T.~Dannert, A.~Marek, and M.~Rampp.
\newblock Porting large {HPC} applications to {GPU} clusters: The codes {GENE}
  and {VERTEX}.
\newblock this conference volume, 2013.

\end{thebibliography}
%%\begin{thebibliography}{99}
%%
%%%%% IOS examples
%%%%\bibitem{r1}
%%%%\textit{Scientific Style and Format: The CBE manual for authors,
%%%%editors and publishers}. Style Manual Committee, Council of Biology Editors.
%%%%Sixth ed. Cambridge University Press, 1994.
%%%%
%%%%\bibitem{r2}
%%%%L.U. Ante, Cem surgere: Surgite postquam sederitis, qui manducatis panem doloris,
%%%%\textit{Omnes} \textbf{13} (1916), 114--119.
%%%%
%%%%\bibitem{r3}
%%%%T.X. Confortavit, \textit{Seras}, Portarum, New York, 1995.
%%%%
%%%%\bibitem{r4}
%%%%P.A. Deus, Ater hoc et filius et mater praestet nobis,
%%%%\textit{Paterhoc} \textbf{66} (1993), 856--890.
%%
%%\end{thebibliography}
\end{document}